\documentclass[runningheads]{llncs}
\usepackage{graphicx}
\usepackage{gensymb}
\usepackage{amssymb}
\usepackage{mathtools}
\usepackage{amsmath}

\DeclarePairedDelimiter{\floor}{\lfloor}{\rfloor}

\begin{document}
\title{A Projection-Based K-space Transformer Network for Undersampled Radial MRI Reconstruction with Limited Training Subjects}
%
\titlerunning{Projection-Based K-space Transformer}
%
\author{Chang Gao\inst{1} \and
Shu-Fu Shih\inst{1} \and
J. Paul Finn\inst{1} \and
Xiaodong Zhong\inst{2}}
%
\authorrunning{C. Gao et al.}
%
\institute{University of California Los Angeles, Los Angeles, CA, United States\\
\email{gaoc@ucla.edu} \and
MR R\&D Collaborations, Siemens Medical Solutions USA, Inc., Los Angeles, CA, United States}
\maketitle 
\begin{abstract}
The recent development of deep learning combined with compressed sensing enables fast reconstruction of undersampled MR images and has achieved state-of-the-art performance for Cartesian k-space trajectories. However, non-Cartesian trajectories such as the radial trajectory need to be transformed onto a Cartesian grid in each iteration of the network training, slowing down the training process significantly.  Multiple iterations of nonuniform Fourier transformation in the networks offset the advantage of fast inference inherent in deep learning. Current approaches typically either work on image-to-image networks or grid the non-Cartesian trajectories before the network training to avoid the repeated gridding process. However, the image-to-image networks cannot ensure the k-space data consistency in the reconstructed images and the pre-processing of non-Cartesian k-space leads to gridding errors which cannot be compensated by the network training. Inspired by the Transformer network to handle long-range dependencies in sequence transduction tasks, we propose to rearrange the radial spokes to sequential data based on the chronological order of acquisition and use the Transformer network to predict unacquired radial spokes from the acquired data. We propose novel data augmentation methods to generate a large amount of training data from a limited number of subjects. The network can be applied to different anatomical structures. Experimental results show superior performance of the proposed framework compared to state-of-the-art deep neural networks.

\end{abstract}
\section{Introduction}
Magnetic Resonance Imaging (MRI) has been widely used for non-invasive diagnosis due to its high resolution and excellent soft tissue contrast. However, its long acquisition time is a drawback and predisposes to artifacts due to patient movement. Undersampling is an efficient way to shorten the acquisition time but the undersampling process itself may produce artifacts and degrade image quality. In the past decades, methods such as compressed sensing (CS) use sparsity constraints to reconstruct images \cite{lustig2008compressed,feng2014golden,feng2016xd} but CS typically suffers from long computational time. The recent development of deep learning combined with CS acquisition schemes enabled fast acquisition and fast reconstruction and achieved state-of-the-art performance on Cartesian trajectories for MRI \cite{yu2017deep,yang2017dagan,quan2018compressed}.

Enforcing k-space data consistency in deep learning-based CS reconstruction is straightforward and time-efficient for Cartesian trajectories, which only needs forward and inverse Fourier transforms between k-space and image space. However, for non-Cartesian trajectories, the non-Cartesian k-space needs to be gridded onto a Cartesian basis prior to Fourier transformation. Gridding is time-consuming and typically needs to be performed with each iteration of the network training and testing, which slows down the process profoundly and offsets the deep learning advantage of fast inference \cite{kofler2021end}. For this reason, a previous study employed a convolutional neural network (CNN) component with very few parameters and implemented a pretraining and fine-tuning strategy to train a radial reconstruction network \cite{kofler2021end}, imposing limits on the application and performance of deep networks. Other work either used image-to-image networks \cite{hauptmann2019real,kofler2019spatio,nezafat2020deep,fan2020rapid,shen2021rapid,chen2021ground} or gridded the radial k-space before the network training \cite{malave2020reconstruction,terpstra2020deep,el2021multi}. For image-to-image networks, the generated images may lack k-space data consistency and therefore compromise image fidelity \cite{kofler2021end}. Gridding non-Cartesian k-space before the network training can avoid the repeated gridding problem, but the network cannot learn to compensate for the gridding error \cite{fessler2007nufft}. In addition, the artifacts due to radial k-space undersampling manifest in the images as global streaks which are hard to remove in the image domain with local convolution kernels.

Inspired by the recent development of the Transformer network for handling long-range dependencies in sequence transduction tasks, we propose to use a projection-based k-space Transformer network to predict unacquired radial spokes from the acquired data. The Transformer network \cite{vaswani2017attention} has outperformed recurrent and convolutional neural networks in language translation \cite{vaswani2017attention}, image generation \cite{parmar2018image} and image classification \cite{dosovitskiy2020image}. Previous works on vision Transformers proposed to crop the image and reshape the image patches to sequential vectors where each vector is a token \cite{dosovitskiy2020image}. However, cropping and reshaping the image will cut off the global streaking artifacts and make the network less effective at removing them.

Similar to the radial GRAPPA work \cite{seiberlich2011improved}, we aim to fill the missing radial spokes using the information from acquired ones. However, radial GRAPPA used local reconstruction kernels and a k-space center calibration region to utilize the multi-coil information. It cannot utilize the global k-space information and only works with radial spokes that are separated by the same angle. We propose to rearrange the radial k-space to sequential data based on the chronological order of acquisition and use the attention mechanism to learn the dependencies between the radial spokes.

To our knowledge, this work was the first to translate MRI k-space into sequential time-series data and to use the Transformer network for MRI raw data prediction and image reconstruction. We used the golden-angle stack-of-radial trajectory as an example to show the projection-based k-space Transformer network and its performance. Novel data augmentation methods were used to generate a large amount of training data from a limited number of subjects. The proposed network was generalized to test on various anatomical structures including the abdominal and pelvic regions.

\section{Projection-based K-space Transformer (PKT)}
\begin{figure}[!t]
\includegraphics[width=\textwidth]{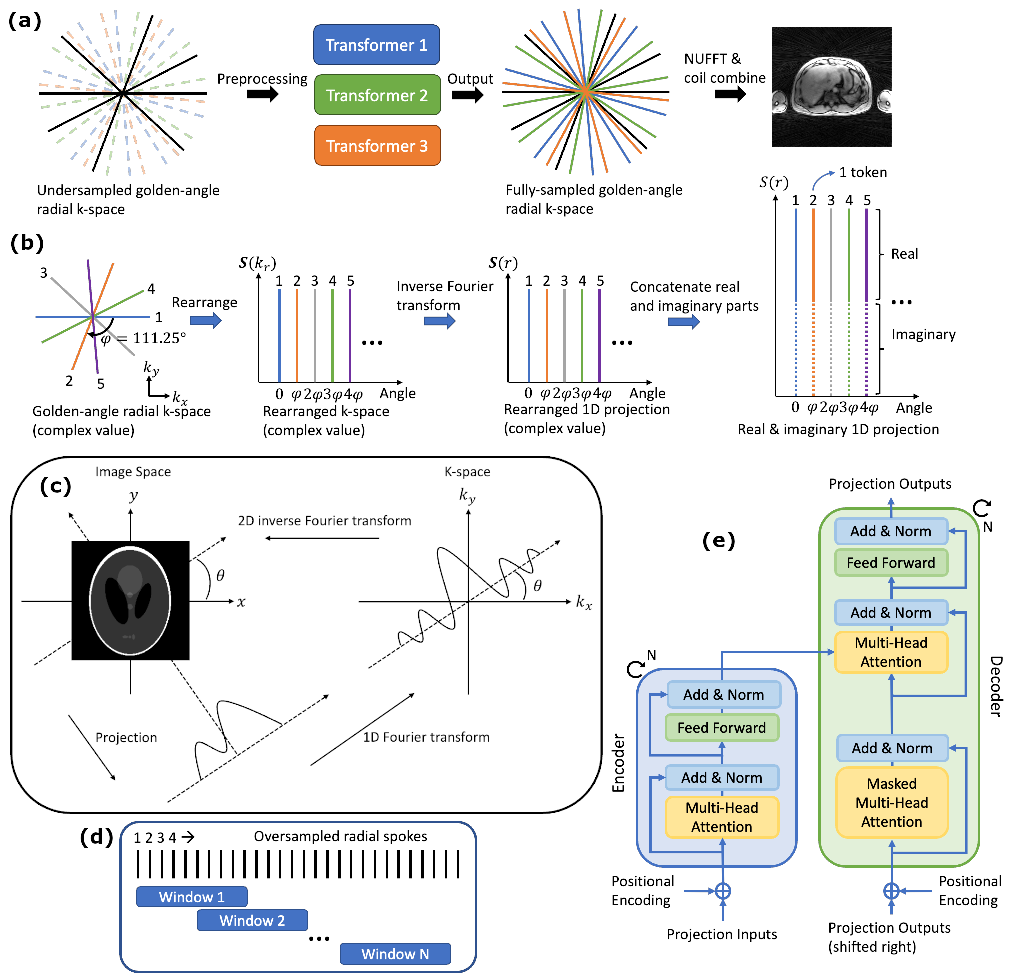}
\caption{(a) Overview of the proposed projection-based k-space transformer (PKT) network. (b) Illustration of the pre-processing of radial k-space. (c) The relationship between the image, projection and k-space. (d) Data augmentation method using over-sampling and sliding window approaches. (e) The Transformer network architecture.} \label{fig1}
\end{figure}

To demonstrate the concept of the proposed framework, a golden-angle-ordered radial k-space is shown as an example in Fig.~\ref{fig1}(a). Starting from 0\degree, each radial k-space spoke is rotated by a golden angle, which is approximately 111.25\degree. The golden-angle acquisition ensures uniform coverage of the k-space for any arbitrary number of consecutive spokes so any newly acquired radial spoke is interleaved with the acquired ones. The spokes that are physically close (in azimuth angle) are dependent on each other. We hypothesize that the dependencies between each of the spokes can be learned by the attention modules in a Transformer network to predict unacquired spokes. We propose to rearrange the radial spokes based on the chronological order of acquisition and encode the temporal spoke index in the network. Note that the same concept can be extended to other non-Cartesian k-space trajectories such as the spiral trajectory or uniform-angle ordering.

\subsection{The Overall Framework}
Suppose there are 256 sampling points along each spoke, a fully-sampled radial k-space needs $256*\pi/2 \approx 402$ spokes according to the Nyquist sampling requirement \cite{feng2014golden}. In this work we aim to achieve 4 times undersampling and thus set the number of spokes of the input to be 100 and the output has 300 new spokes to supplement the acquired ones. We propose to train 3 independent Transformer networks, each predicting 100 consecutive spokes of different angles.

As illustrated in Fig.~\ref{fig1}(a), the undersampled radial k-space has 100 spokes $(p_i,p_{i+1},...,p_{i+99})$ (in black), where $i$ is the spoke index of the first spoke in each sequential set of spokes. Transformers 1, 2 and 3 take the pre-processed radial spokes as input and predict unacquired k-space radial spokes independently and in parallel. Spokes $(p_{i+100},p_{i+101},...,p_{i+199})$ (in blue), $(p_{i+200},p_{i+201},...,p_{i+299})$ (in green) and $(p_{i+300},p_{i+301},...,p_{i+399})$ (in orange) are the outputs of Transformers 1, 2 and 3, respectively. The corresponding spoke indices are encoded in the Transformer network to distinguish spokes at different angles. The input and 3 outputs are combined in the order of the spoke index to form a fully-sampled radial k-space. To reconstruct the final images, the radial k-space is nonuniformly fast Fourier transformed (NUFFT) using the Bart toolbox \cite{uecker2016bart} and the images from different coils are combined using adaptive coil combination \cite{walsh2000adaptive} using MATLAB R2021b (MathWorks).

Fig.~\ref{fig1}(b) shows pre-processing steps of radial k-space. For each single coil 2D slice, the golden-angle-ordered radial spokes are rearranged based on the chronological order of acquisition. Then each of the k-space spokes is inverse Fourier transformed to generate a projection vector at each angle. Fig.~\ref{fig1}(c) shows the relationship between the image, projection and k-space. The projection signals span the whole image compared to the centrally condensed k-space signals, so the network can better learn the image details and the reconstructed image is less sensitive to network errors. The real and imaginary parts of each complex-valued projection are concatenated to a single long vector prior to input to the network. Because the frequency encoding line ($k_x$) is typically 2 times oversampled, the complex-valued projection vector is of length 512 and the concatenated projection vector is of length 1024. The input and output are of size 1024x100 with each projection vector as a token in the network.

\subsection{Data Augmentation}
Generally, large training data are necessary to train deep learning networks, which can pose practical challenges. With the proposed k-space Transformer framework, several data augmentation methods were used to enlarge the training data size with limited subjects. The data augmentation methods used in this work include (1) acquisition from multiple anatomical regions, (2) over-sampled 3D multi-coil acquisition and (3) a sliding window approach to generating more training data. Scanning from multiple anatomical regions not only enlarges the data size but also increases the data diversity and can help the generalization of the network on various anatomical images. Each 3D multi-coil acquisition is first inverse fast Fourier transformed (FFT) along $k_z$ to generate multi-coil multi-slice 2D radial k-space. Because each slice has over-sampled k-space data, a sliding window approach was used to generate more training data from the same single-coil slice, as illustrated in Fig.~\ref{fig1}(d). Each window contains 400 spokes and the step size is 200 spokes. Suppose $n_{sub}$ subjects are acquired for training, $n_{reg}$ anatomical regions for each subject, $n_{slc}$ slices and $n_{coil}$ coils for each 3D acquisition, and $M$ spokes for each slice, the effective training data size $DS$ will be:
\begin{equation}
    DS = n_{sub} \times n_{reg}\times n_{slc} \times n_{coil} \times \floor{M/400}
\end{equation}
where $\floor{}$ represents the floor function.

\subsection{Transformer Network}
Considering the relatively long length of sequential radial spokes, the application of the Transformer network is expected to learn the long-range dependencies between the spokes. As shown in Fig.~\ref{fig1}(e), the Transformer has an encoder-decoder structure. The encoder maps an input of the acquired radial spokes to sequential representations, which was used by the decoder to generate the sequential output. Each spoke is treated as a token in the network. At each step the decoder is auto-regressive, using the previously generated spokes to generate the next single spoke. The encoder consists of $N=6$ identical stacks, each having a multi-head self-attention sub-layer and a fully connected feed-forward sub-layer. After each sub-layer, a residual connection \cite{he2016deep} is implemented followed by layer normalization \cite{ba2016layer}. The decoder also has $N=6$ identical stacks. In addition to the two sub-layers in encoder, the decoder has a third sub-layer, which learns the multi-head attention from the outputs. The output is offset by one position and is masked for each iteration so that the prediction for position $pos$ depends only on the known outputs at positions less than $pos$.

The angle information of the radial spokes is highly correlated with their dependency. Therefore, positional encoding is implemented for the input and output to make use of the angle information. For a set of consecutive spokes $(k_1, ..., k_n)$ each having a length of $d_{model}$, the positional encoding is calculated using sine and cosine functions:
\begin{equation}
\begin{aligned}
    PE_{i_s, 2j} = sin(i_s/10000^{2j/d_{model}}) \\
    PE_{i_s, 2j+1} = cos(i_s/10000^{2j/d_{model}})
\end{aligned}
\end{equation}
where $i_s$ is the relative spoke index and $i_s \in \{0,1,...,399\}$ and $j$ is the dimension. The positional encoding is directly added to the input and output. Using this positional encoding strategy, consecutive input and output are not necessary which allows more flexibility for the network design. For example, when the input is $(k_1, ..., k_n)$, it is not necessary for the output to be $(k_{n+1}, ..., k_{2n})$ only, where $(k_m, ..., k_{m+n-1})$ is also feasible ($m>n$). This strategy allows to train independent Transformers to predict spokes at different angles and combine all predictions and the input for the final reconstruction.

\section{Experiments and Results}
\subsection{Dataset}
The study was compliant with the Health Insurance Portability and Accountability Act and approved by the Institutional Review Board. With written informed consent from each subject, we acquired radial MRI data using a prototype golden-angle, gradient-echo stack-of-radial sequence from 17 healthy subjects in the abdomen, pelvis, thigh and leg (16 subjects on a 3T MAGNETOM Prisma\textsuperscript{fit} and 1 subject on a 3T MAGNETOM Skyra, Siemens Healthcare, Erlangen, Germany). The sequence parameters were: 512 sampling points each spoke (with 2x oversampling along frequency encoding direction), 3500 spokes per partition and 32 partitions. For each subject, images were acquired at 3 regions on average and each acquisition had around 30 coils. The reference images were acquired with oversampling with 3500 radial spokes and were retrospectively downsampled to generate undersampled images.

We divided the data into training (8 subjects), validation (2 subjects) and testing (7 subjects) datasets. Because of the data augmentation strategies described in Section 2.2, although the number of subjects was relatively small in this study, the generated data size was considerably large, comprising 193536 training data, 61440 validation data and 11968 testing data.

\subsection{Implementation Details}
We used Mean-Squared-Error (MSE) loss, i.e. $l_2$ loss, to minimize the distance between the output and the ground truth:
\begin{equation}
    \min_{\theta}L_{l_2}(\theta) = ||T_{\theta}(x)-y||^2_2
\end{equation}
where $\theta$ represents the network parameters, $T$ represents the Transformer, $x$ is the input and $y$ is the ground truth. Implementation and training of the proposed network were completed in Python 3.7.7 (Python Software Foundation) using the Pytorch library 1.8.1 \cite{paszke2017automatic} on a computer with three NVIDIA Quadro RTX 8000 GPUs. The networks were trained for 100 epochs with a batch size of 400. An Adam optimizer with $\beta_1=0.9$, $\beta_2=0.98$, and $\epsilon=10^{-9}$ was used. The hyperparameters were $d_{model}=1024$, 16 attention heads, dimensions of keys and values $d_k=64$, $d_v=64$, dropout=0.1. The total training time was around 3 days. The reference data containing 400 spokes was regarded as the full k-space acquisition and was used to reconstruct the reference standard image.

\subsection{Performance Evaluation}
For quantitative evaluation, we used Normalized MSE (NMSE), peak signal-to-noise ratio (PSNR) and structural similarity index (SSIM) between the reconstructed images and the reference images. NMSE was also calculated for the projection data. All images were normalized by the 90th percentile before the assessment. Statistical differences between the proposed PKT and the baseline methods were assessed using a one-way ANOVA model \cite{howell2012statistical} with the reconstruction method being the independent variable.

As representative state-of-the-art techniques, we also trained image-based U-Net \cite{ronneberger2015u}, DenseUnet \cite{xiang2018deep} and SEUnet \cite{hu2018squeeze} to compare with our proposed PKT network. DenseUnet used dense connections between U-Net layers and SEUnet added channel-wise attention modules to the U-Net. We reconstructed 2D images with 100 and 400 spokes as the input and output images, respectively. All network training were completed in Python 3.7.7 using the Pytorch library 1.8.1 on a computer with a NVIDIA V100 GPU. The initial learning rate was 0.0001 which was reduced with a factor of 0.7 if no improvement shown in 200 iterations. The training was stopped when the learning rate was below $1\times10^{-6}$ or the max epoch number reached 100.

\subsection{Results}
\begin{figure}[!t]
    \centering
    \includegraphics[width=\textwidth]{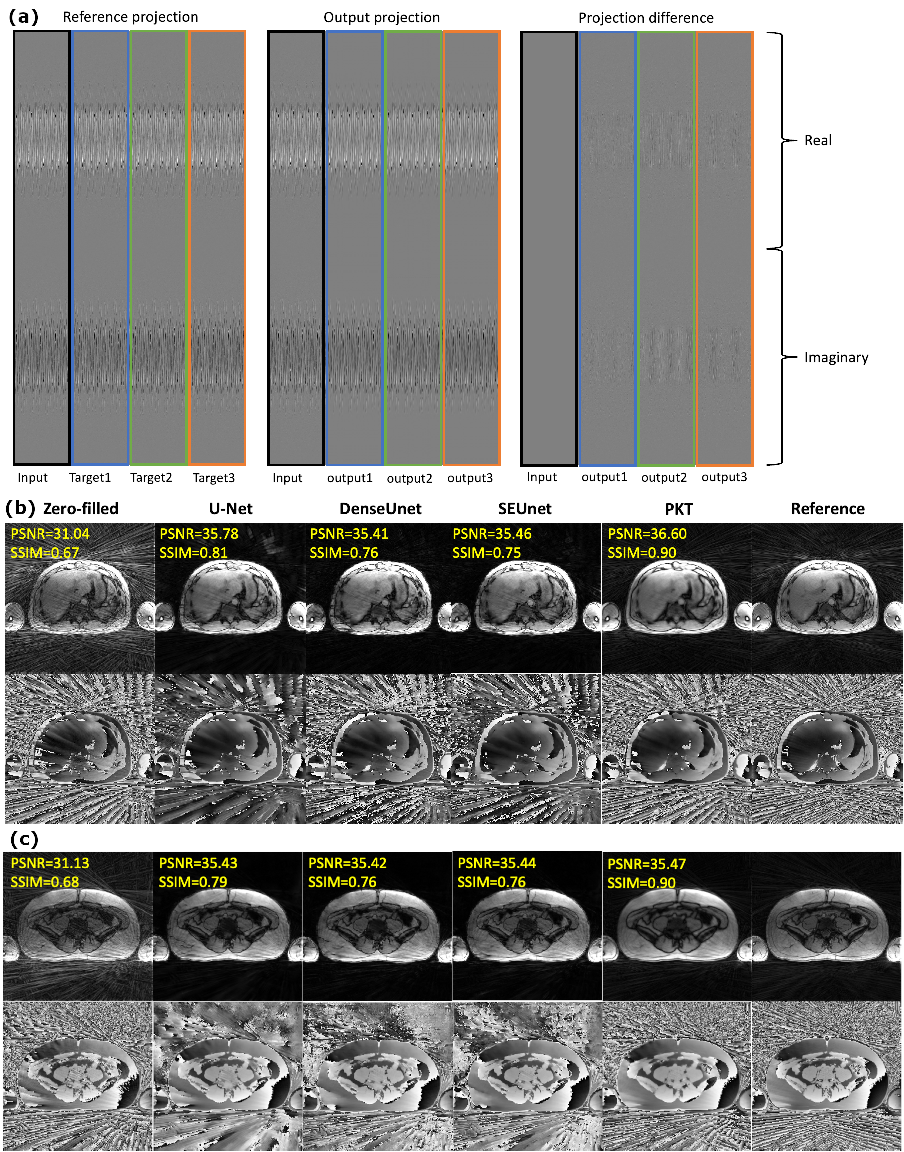}
    \caption{(a) A representative comparison of the network outputs and the reference k-space projection data. (b) and (c) show representative images of the proposed PKT compared with baseline reconstructions along with the reference. Fig. (b) shows an abdominal example and Fig. (c) shows a pelvis example. In both Fig. (b) and (c), the top row shows the magnitude images and the bottom row shows the phase images.} \label{fig2}
\end{figure}

The network generated k-space projection data is shown in Fig.~\ref{fig2} (a). Representative abdominal and pelvis images of the proposed PKT compared with baseline reconstructions and the reference images are shown in Fig.~\ref{fig2} (b) and (c), respectively. The proposed PKT successfully removed the streaking artifacts and improved signal-to-noise in the magnitude and phase abdominal images compared to the other reconstructions. Fig.~\ref{fig2} (c) shows that the PKT reconstruction can also be applied to the pelvis region to reduce streaking artifacts and noise.

The NMSE of all testing projection data is 0.0385±0.0497. Quantitative evaluations of the proposed PKT network compared with baseline reconstructions with respect to the reference images are shown in Table~\ref{table1}. The proposed PKT have a significantly lower NMSE ($p<.0001$) and significantly higher PSNR ($p<.0001$) and SSIM ($p<.0001$) compared to all other reconstructions. The results indicate that the proposed PKT reconstructed images had fewer streaking artifacts and less noise compared to the other methods.

\begin{table}
\caption{Quantitative comparisons of the proposed PKT with baseline reconstructions. Values are reported as mean $\pm$ std across test subjects.}\label{table1}
\centering
\setlength{\tabcolsep}{12pt}
\begin{tabular}{l|l|l|l}
\hline
Methods & NMSE & PSNR & SSIM (\%)\\
\hline
Zero-filled & 0.1389$\pm$0.0275 & 27.28$\pm$2.10 & 59.19$\pm$5.25\\
U-Net & 0.0755$\pm$0.0527 & 30.44$\pm$2.87 & 71.81$\pm$6.23\\
DenseUnet & 0.0765$\pm$0.0443 & 30.28$\pm$2.87 & 67.20$\pm$6.16\\
SEUnet & 0.0759$\pm$0.0447 & 30.34$\pm$2.89 & 67.54$\pm$6.45\\
{\bfseries PKT} & {\bfseries 0.0456$\pm$0.0101} & {\bfseries 32.13$\pm$2.26} & {\bfseries 83.44$\pm$4.84}\\
\hline
\end{tabular}
\end{table}

\section{Conclusion}
In this study, we propose to rearrange MRI radial projection to sequential time-series data and predict unacquired radial spokes using projection-based k-space Transformer (PKT) networks. We demonstrated PKT using golden-angle-ordered 3D radial data. Results showed PKT removed streaking artifacts and noise and had significantly better performance than other state-of-the-art deep learning neural networks. The novel data augmentation strategies in this study generated sufficient training data from a limited number of subjects by acquiring over-sampled multi-coil 3D data at various anatomical regions with the sliding window approach. We demonstrated that PKT could be applied to various anatomical structures including the abdominal and pelvic regions.

%
%
%
\bibliographystyle{splncs04_unsorted}
\bibliography{references.bib}
%




\end{document}